\documentclass[twocolumn,showpacs,preprintnumbers]{revtex4}
\usepackage{graphicx}
\usepackage{dcolumn}
\usepackage{bm}

\input{tcilatex}

\begin{document}

\title{Pressure effects in Pr\textit{T}$_{2} $B$_{2}$C (\textit{T} = Co, Ni,
Pt): Applied and chemical pressure}
\author{R Falconi}
\author{A Dur\'{a}n$^{1}$}
\author{M N\'{u}\~{n}ez-Regueiro$^{2}$}
\author{R Escudero$^{3}$}
\affiliation{Divisi\'{o}n Acad\'{e}mica de Ciencias B\'{a}sicas,\linebreak Universidad Ju%
\'{a}rez Aut\'{o}noma de Tabasco. Cunduacan, Tabasco. 86690 A. Postal 24. M%
\'{E}XICO}
\affiliation{$^{1}$Centro de Nanociencias y Nanotecnolog\'{\i}a,\linebreak Universidad
Nacional Aut\'{o}noma de M\'{e}xico. A. Postal 2681, Ensenada, Baja
California 22800, M\'{E}XICO}
\affiliation{$^{2}$Institut N\'{e}el, Centre Nationale de la Recherche Scientifique \&
Universit\'{e} Joseph Fourier, BP 166 38042 GRENOBLE, FRANCE.}
\affiliation{$^{3}$Instituto de Investigaciones en Materiales, Universidad Nacional Aut%
\'{o}noma de M\'{e}xico. A. Postal 70-360. M\'{e}xico, D.F. 04510 M\'{E}XICO.}
\date{\today }

\begin{abstract}
High-pressure electrical resistivity $\rho _{ab}$ measurements on\
intermetallic Pr(Co, Ni, Pt)$_{2}$B$_{2}$C compounds were performed down to
2K. \ At room pressure the $\rho _{ab}(T)$ curves for the non
superconducting Pr(Co, Ni)$_{2}$B$_{2}$C compounds exhibit magnetic
correlations at about 10 and 4 K, respectively. At low temperatures, PrCo$%
_{2}$B$_{2}$C shows a large spin-dependent electron scattering in comparison
to PrNi$_{2}$B$_{2}$C. Under applied pressure the magnetic scattering tends
to be suppressed more effectively in PrCo$_{2}$B$_{2}$C than in PrNi$_{2}$B$%
_{2}$C. The low temperature behavior of $\rho _{ab}(T,P)$ for\ PrNi$_{2}$B$%
_{2}$C and PrCo$_{2}$B$_{2}$C suggests a spin fluctuations mechanism. In the
other hand PrPt$_{2}$B$_{2}$C compound shows superconductivity at about 6 K
and under pressure its superconducting transition temperature tends to be
degraded at a rate dT$_{C}$/dP = -0.34\ K/GPa, as expected in compounds with
transition metals. The experimental results in Co, Ni and Pt based compounds
are analyzed from the point of view of the external and chemical internal
pressure effects.
\end{abstract}

\maketitle

\section{Introduction}

The phenomenon of superconductivity appearing in compounds having magnetic
elements, has received noticeable attention during the last three decades
due to the great variety of exotic electronic and magnetic correlations it
involves. Particularly interesting for this topic, has been the discovery of
the quaternary intermetallic compounds; \textit{R}Ni$_{2}$B$_{2}$C, (\textit{%
R} = rare earths, Y, Sc, Th) \cite{Nagaraj,Cava1} where coexistence of
antiferromagnetism and superconductivity has been observed as for example in 
\textit{R} = (Tm, Er, Ho, Dy, Lu) \cite{Cava1,Cava2,Tomy,Cho,Muller,Eisaki}.
As far as we know several \textit{R}T$_{2}$B$_{2}$C intermetallics
compounds, with different rare earths (\textit{R}) and transition metal (T)
combinations, have been synthesized \cite%
{Cava2,Muller,Cava3,Carter,Sampath,Massalami,Gupta2006,Anand2007,Massalami2009}
and most of them, such as HoNi$_{2}$B$_{2}$C \cite{LinPRBHolmioSolo}, show
superconductivity in spite of the presence of the rare earth magnetic
element.

The influence of the transition metal magnetism in the magnetic properties
of these compounds, seem to be of minor importance compared to that of the
rare earth ions, whose magnetic moments apparently impose the magnetic
ordering at all. Thus in some borocarbides with 3\textit{d} transition
elements such as Ni and Co, neutron-diffraction measurements \cite{LynnPRB55}%
, and electronic transport measurements \cite{Massalami,Schmidt}, have
revealed that no significant magnetic moment develops in the T sites. Local
structure studies at the Ni site, using Mossbauer spectroscopy on $^{57}$Fe
doped (1 at \%) samples, also support this fact \cite{Sanchez}.
Interestingly, the T elements play an indirect role in the magnetism of
magnetic \textit{R}T$_{2}$B$_{2}$C systems through the spatial dependent
indirect RKKY exchange interactions \cite{Eisaki,Gratz}, that govern the
magnetic ordering in these compounds. On the other hand, the electronic
influence of the T elements in the superconductivity of \textit{R}T$_{2}$B$%
_{2}$C is more relevant than that for the \textit{R} , B and C elements.
This is particularly true for the Ni based borocarbide superconductors,
where the density of state at the Fermi level is mainly due to the Ni 
\textit{3d} bands \cite{Loureiro}. On the contrary, in a comparative study
of the structure and superconducting properties of \textit{R}Ni$_{2}$B$_{2}$%
C, Loureiro \textit{et al. }\cite{Loureiro}, showed that the superconducting
state is more strongly affected by the magnetism of the \textit{R} ion than
for the \textit{R}-ion size, at least for \textit{R} between Dy and Tm.
However, the role of the magnetism and ion-size of T elements in the
superconductivity of \textit{R}T$_{2}$B$_{2}$C when magnetic \textit{R} ions
are presents, is not clear yet. In this work we aboard the study of a
particularly interesting case: the PrT$_{2}$B$_{2}$C compounds, with T = Ni,
Co, and Pt which have revealed many peculiarities. PrNi$_{2}$B$_{2}$C, and
PrCo$_{2}$B$_{2}$C do not superconduct as measured down to 0.3 K \cite%
{Narozhnyi}, however, PrPt$_{2}$B$_{2}$C does superconduct at 6 K, even in
the magnetic Pr$^{+3}$ ion presence \cite{Cava4,Dhar}. Noticeably, PrPt$_{2}$%
B$_{2}$C does not show any magnetic ordering at low temperatures \cite{Dhar}
but in contrast, Pr-Ni and Pr-Co based borocarbides, develop
antiferromagnetic ordering at about 4 and 8.5 K, respectively \cite%
{Duran1PRB,Duran2PRB}. Recently, magnetoresistance and specific heat studies
in Pr(Co, Pt)$_{2}$B$_{2}$C \cite{Duran2PRB,Morales} have pointed out that
spin fluctuation mechanism is involved in the electronic behavior of these
two compounds. However, although evidence for spin fluctuations can be
deduced from certain features in the electronic transport measurements, the
interpretation of those properties could be not so clear. High pressure
experiments in spin fluctuators such as RCo$_{2}$ \cite{HauserRCo2pressure},
CeNi$_{5}$ \cite{JMMMCeNi5spinFlucPres}, and UPt$_{3}$ \cite%
{BrodaleUPt3pressure} have proven to be a useful tool in order to make clear
if a spin fluctuation mechanism is occurring in such systems. The aim of
this paper is to enlighten the influence of the chemical and external
applied pressure on the superconducting state and magnetic scattering at low
temperature for the three Pr-based borocarbides: Pr(Ni, Co, Pt)$_{2}$B$_{2}$%
C. We analyzed the changes of the resistivity as a function of pressure and
temperature. We assume that interactions between itinerant electrons plays
an important role in the low temperature resistivity characteristic, and
those can be modified by applied external or internal chemical pressure.

\section{Experimental details}

Three compounds were prepared: samples of PrCo$_{2}$B$_{2}$C, PrNi$_{2}$B$%
_{2}$C, and PrPt$_{2}$B$_{2}$C. The single crystals were grown by cold
copper crucible method as described by Dur\'{a}{}n \textit{et al} \cite%
{Duran2PRB}. All samples were characterized by X-ray diffraction using a
Bruker P4 diffractometer, with monochromatized Mo-K$\alpha $ radiation. The
cell parameters were: \textit{a }$=3.6156$ \AA , and \textit{c} $=10.3507$ 
\AA \thinspace\ for PrCo$_{2}$B$_{2}$C,\ \textit{a} $=3.6996$ \AA , and 
\textit{c} $=9.9885$ \AA\ for PrNi$_{2}$B$_{2}$C, and \textit{a }$=3.8373$ 
\AA , \textit{c} $=10.761$ \AA\ for PrPt$_{2}$B$_{2}$C samples. Resistivity
measurements in the \textit{a-b} plane were performed by the four-probe
technique using gold wires of $10$ $\mu m$ diameter as electrical contacts.
Pressure experiments were performed by using a micro-cryogenic diamond anvil
cell MCDAC (cell piston-cylinder type Be-Cu cell) consisting of two
diamonds, each of $0.5$ $mm$ culet size. A Cu-Be gasket was preindented and
a $150$ $\mu m$ diameter hole was drilled at the center. The samples used
have dimensions of about $80\times 15\times 40$ $\mu m^{3}$ and were placed
in the gasket hole. The transmitting pressure medium was MgO powder. The
metallic gasket was electrically insulated, pressing over it Al$_{2}$O$_{3}$
powder of $1$ $\mu m$ grain size. As the MCDAC pressure increases, the wires
used to measure the electrical resistance may be cut off at the edge of the
diamonds because of the diamond indentation. To reduce this problem, we used
a thin aluminum foil placed under the four gold wires, with this set up
frequently we reached high quasi-hydrostatic pressures in the range about 6
GPa. Also, in order to prevent motion of the sample and of the electrical
leads, at the initial compressing, a thin mylar film was placed over them.
Additional pressure experiments in polycrystalline Pr(Ni, Pt)$_{2}$B$_{2}$C
compounds up to about 21 GPa were made using a sintered diamond Bridgman
anvil apparatus, with a pyrophyllite gasket and two steatite disks as the
pressure medium. For determination of the pressures a Pb manometer were used.%
\FRAME{ftbpFU}{3.269in}{3.7446in}{0pt}{\Qcb{(Color online) Variation of the
T-T shortest length between 3\textit{d} ions for the transition metal T and
PrT$_{2}$B$_{2}$C compounds, with T = (Co, Ni, Pt). It also shows the
behavior of the cell parameters as a function of the T-size.}}{\Qlb{Fig.1}}{%
fig1_falconi.eps}{\special{language "Scientific Word";type
"GRAPHIC";maintain-aspect-ratio TRUE;display "ICON";valid_file "F";width
3.269in;height 3.7446in;depth 0pt;original-width 8.2858in;original-height
9.5008in;cropleft "0";croptop "1";cropright "1";cropbottom "0";filename
'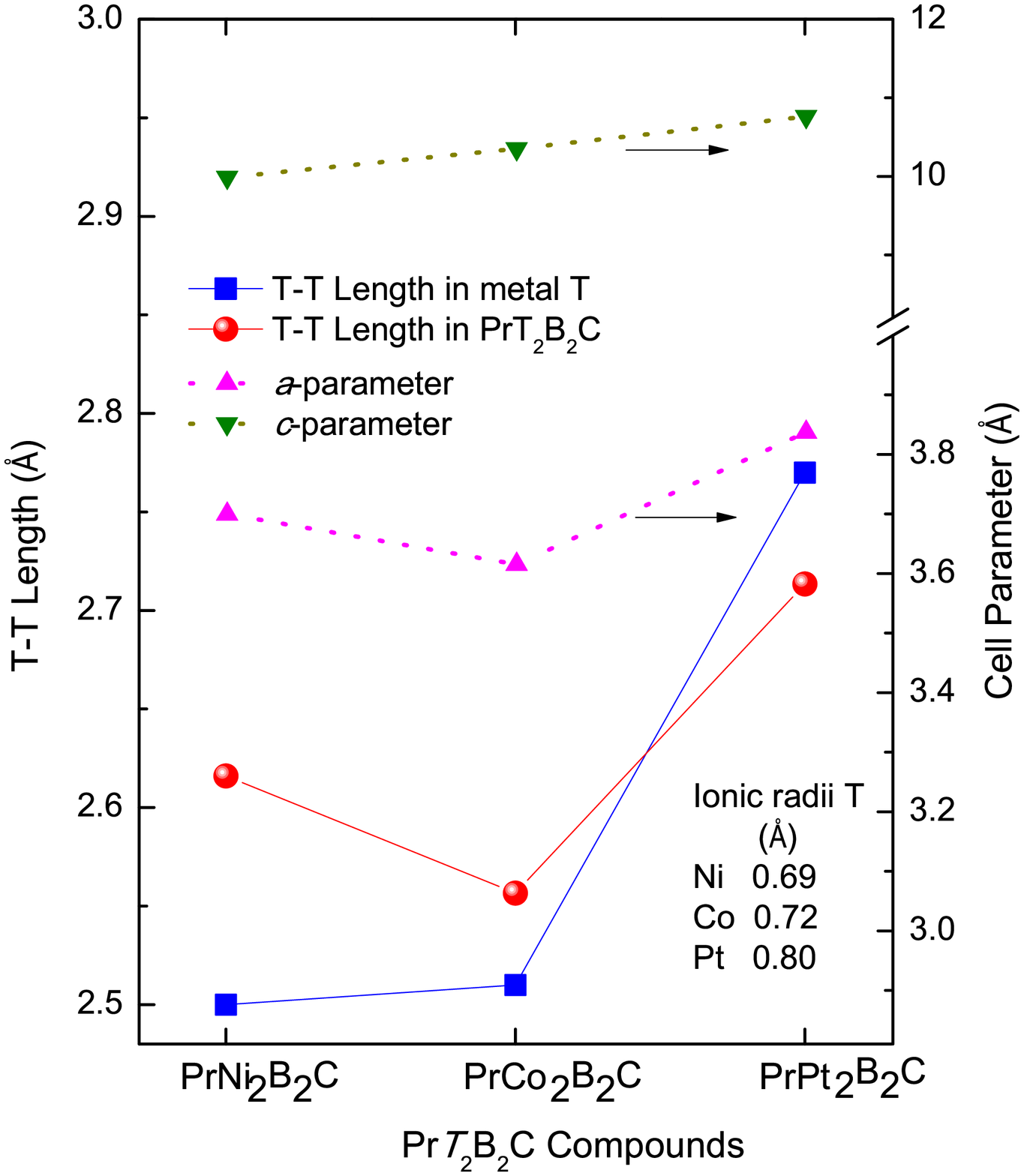';file-properties "XNPEU";}}

\section{Results and discussion}

It is known that the \textit{R}Ni$_{2}$B$_{2}$C compounds crystallize in the
tetragonal body-centered structure (space group \textit{I4}/mmm) and when
the rare earth atom radii is increased (\textit{R} goes from La to Lu), the 
\textit{c}-parameter becomes larger whereas the \textit{a}-parameter
decreases \cite{Siegrist,LynnPRB55}. This structural behavior can be
accounted by for the rigidity of the B-C and Ni-B bonds and the variable
tetrahedral angle in the NiB$_{4}$ unit. Distortions of this tetrahedral
unit are claimed to be a decisive parameter for Tc in non magnetic or
antiferromagnetic \textit{R}Ni$_{2}$B$_{2}$C and \textit{R}NiBC compounds 
\cite{SanchezTetrahedro}. In the case when the size of the transition
element is increased, maintaining the same rare earth element, the
structural behavior of the unit cell seems to be slightly different. Fig. 1
displays the T-T shortest length between 3\textit{d }ions, particularly in
the Pr(Ni, Co, Pt)$_{2}$B$_{2}$C compounds and that for the metal T. Also it
shows the behavior of the \textit{a} and \textit{c} parameters for each
compound. We note that increasing the ionic radii T-size causes an
increasing of the \textit{c}-parameter and an anomaly behavior of the 
\textit{a}-parameter for the PrCo$_{2}$B$_{2}$C compound. At first glance
and according to the figure, this anomaly is correlated with the variation
of the T-T shortest bond in the framework of the PrT$_{2}$B$_{2}$C
structure, and not with the ionic radii size of the T element.\FRAME{ftbpFU}{%
3.3572in}{2.5884in}{0pt}{\Qcb{(Color online) Normalized resistivity at 295 K 
$\protect\rho _{ab}(T)/\protect\rho _{ab}(295$\ K$)$) of PrNi $_{2}$B$_{2}$%
C, PrCo$_{2}$B$_{2}$C, and PrPt$_{2}$B$_{2}$C compounds at room pressure.
The three systems present metallic conductivity from room temperature to
down 25 K. Inset shows the low temperature variation of the normalized
resistivity.}}{\Qlb{Fig2}}{fig2_falconi.eps}{\special{language "Scientific
Word";type "GRAPHIC";maintain-aspect-ratio TRUE;display "ICON";valid_file
"F";width 3.3572in;height 2.5884in;depth 0pt;original-width
10.0024in;original-height 7.7012in;cropleft "0";croptop "1";cropright
"1";cropbottom "0";filename '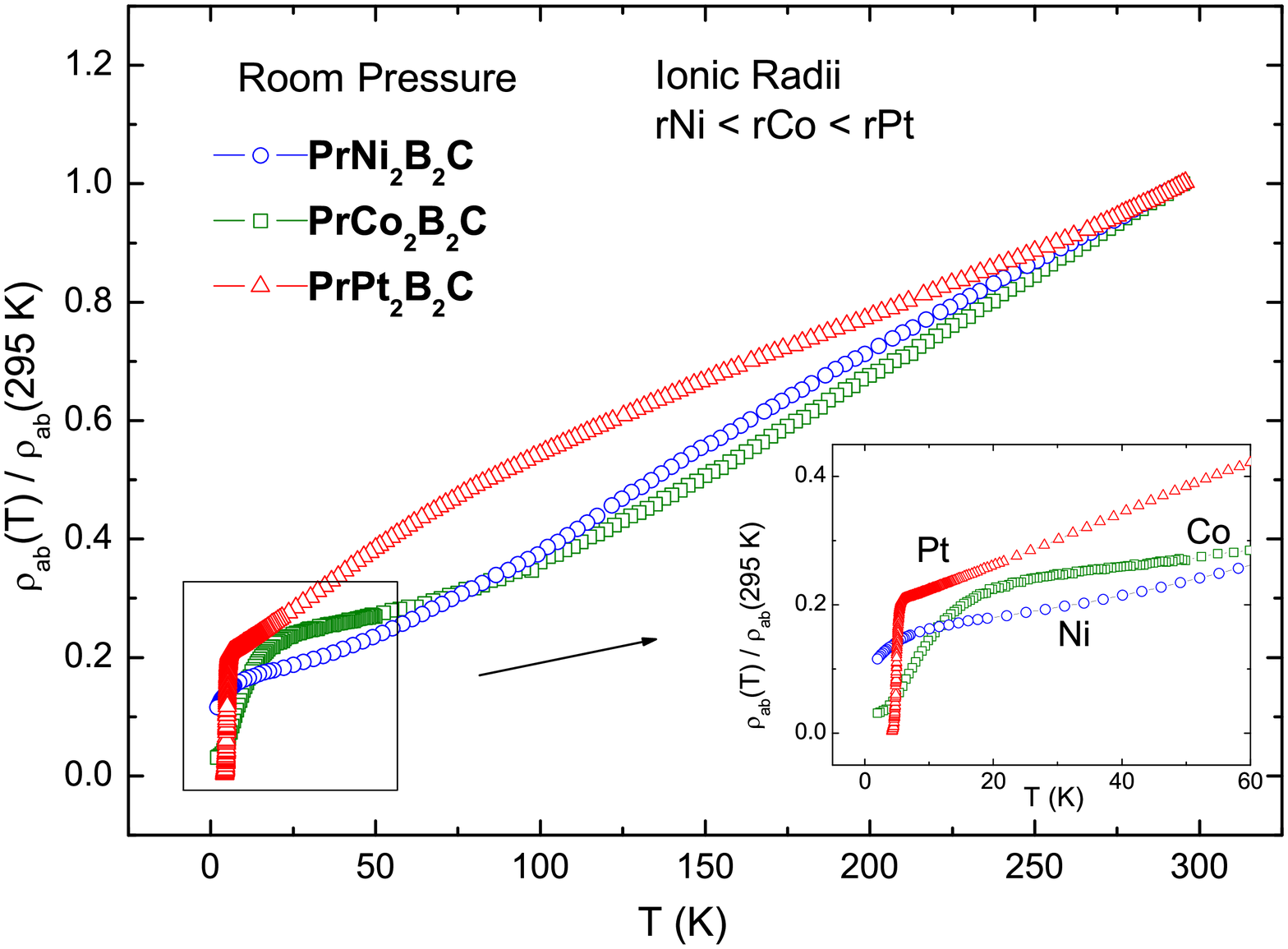';file-properties "XNPEU";}}

Fig. 2 shows the normalized electrical resistivity in the \textit{a-b} plane
as a function of temperature for PrNi$_{2}$B$_{2}$C, PrCo$_{2}$B$_{2}$C, and
PrPt$_{2}$B$_{2}$C single crystals at room pressure. The three compounds
present metallic characteristic. At low temperature Ni and Co based
compounds show notable similarities, but are not superconductors, whereas
PrPt$_{2}$B$_{2}$C has a sharp superconducting temperature at about 6 K. The
residual resistivity ratios RRR for the first two compounds are $9$ and $33$
respectively, whereas for PrPt$_{2}$B$_{2}$C is $5.5$. The residual
resistivity\ $\rho _{0}$ of all three is sample dependent, varying between 5
and 25 $\mu \Omega cm.$ The main panel of Fig. 2 shows some interesting
characteristics: Ni and Co based compounds present a notable positive
curvature from about 150 to 50 K, whereas Pt compound presents a wide bump
from about 250 to 20 K. These notable differences may signal a clearly
distinctive influence of the crystalline field at high temperatures. A
gradual but pronounced drop in resistivity comes to disturb the linear
variation to about 8 and 20 K for PrNi$_{2}$B$_{2}$C, and PrCo$_{2}$B$_{2}$%
C, respectively. Such resistivity behavior at relatively low temperature is
typical for magnetic elements of the \textit{R}Ni$_{2}$B$_{2}$C series, and
it has been associated with a decrease of the magnetic scattering of the
conduction electrons by rare earth ions (\cite{MiprimerPRB} and reference
therein). However, according to the results present below, it is possible
that other mechanism involving conduction electrons could also develop at
low temperature. The case for PrPt$_{2}$B$_{2}$C is quite different, after
following an upward curvature it becomes superconducting at about 6 K.
Magnetic and heat capacity measurements in this compound \cite{Dhar}, have
revealed a nonmagnetic ground state for Pr ions due to CEF effects, which is
claimed to be the reason for superconductivity. The inset of the Fig. 2
shows the resistivity behavior from 60 to 2 K for the three compounds at
room pressure. At first glance, increasing the transition metal radius\
corresponds to a major resistivity droop at low temperature.\FRAME{ftbpFU}{%
3.7109in}{2.9585in}{0pt}{\Qcb{(Color online) Linear fit to $\protect\rho $(T$%
^{2}$) from 2 to 8 K for PrCo$_{{\protect\small 2}}$B$_{{\protect\small 2}}$%
C. The low temperature behavior of $\protect\rho $(T) follows a T$^{2}$\ low
with a cuadratic coefficient A = 0.08 $\protect\mu \Omega cm/K^{2}.$ Inset
shows the $\protect\rho (T)$\ behavior at low temperatures.}}{\Qlb{Fig3}}{%
fig3_falconi.eps}{\special{language "Scientific Word";type
"GRAPHIC";maintain-aspect-ratio TRUE;display "ICON";valid_file "F";width
3.7109in;height 2.9585in;depth 0pt;original-width 6.2881in;original-height
5.0055in;cropleft "0";croptop "1";cropright "1";cropbottom "0";filename
'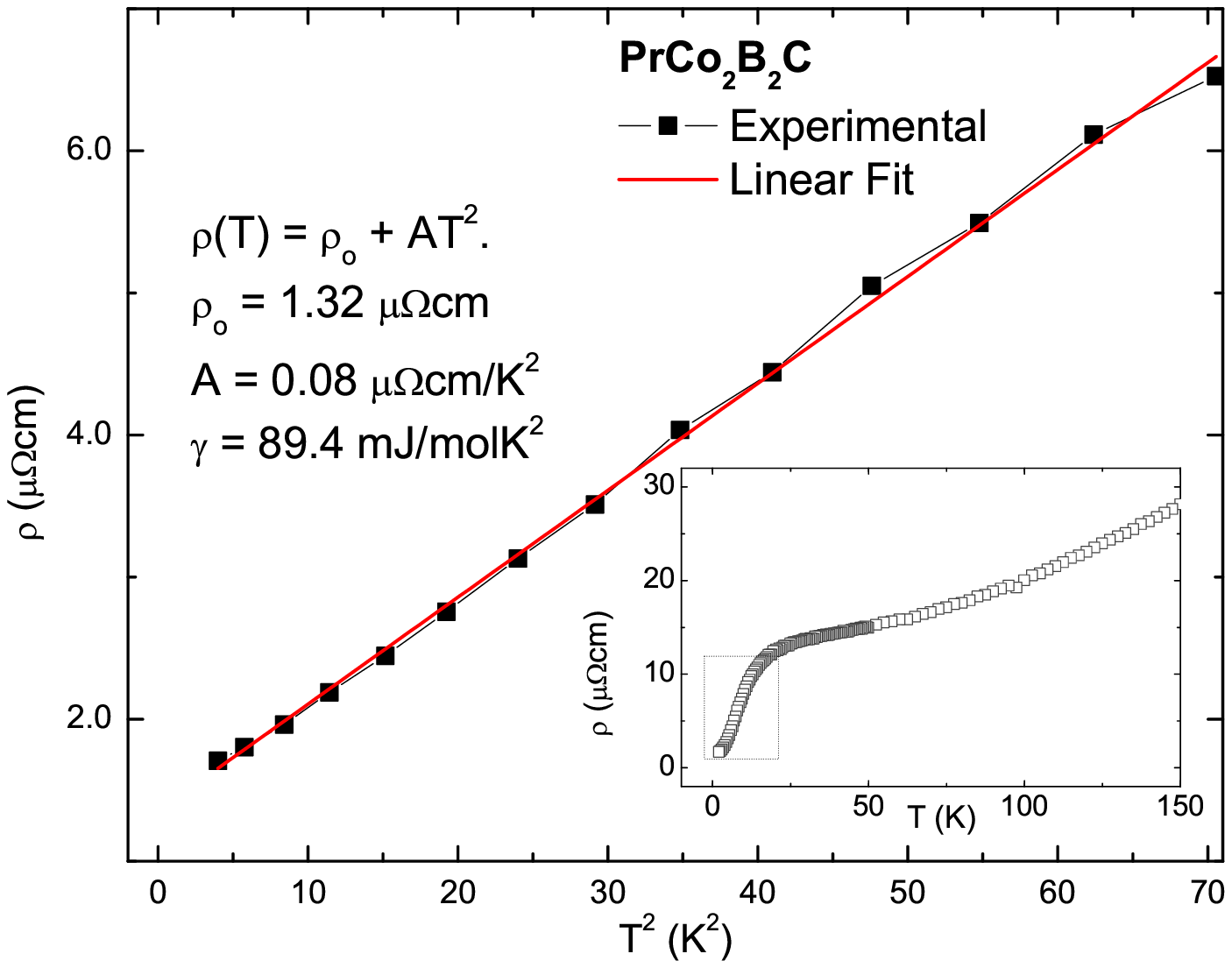';file-properties "XNPEU";}}

The $\rho _{ab}(T)$ curve for PrCo$_{2}$B$_{2}$C from about 2 to 8 K shows a
clear T$^{2}$-law dependence with a cuadratic coefficient A, equal to 0.08 $%
\mu \Omega $cm/K$^{2}$, see Fig. 3. This low-temperature resistivity
behavior is similar to the observed in heavy fermion systems, as for
example; YbNi$_{2}$B$_{2}$C and UPt$_{3}$ compounds \cite{Yatskar,Stewart},
and could be attributed to spin fluctuations \cite{Morales,Moriya}. Thus, in
a similar compound but simpler, RCo$_{2}$, a T$^{2}$ dependence has been
found at low temperatures, which is due to spin fluctuating characteristics 
\cite{HauserRCo2pressure}. The fact that the magnitude of the cuadratic
coefficient \textit{A} of $\rho (T)$ for PrCo$_{2}$B$_{2}$C is of the order
of that for RCo$_{2}$ (\cite{MassalammiJMMM2004}, found this coefficient as
big as three orders of magnitude but in polycrystalline PrCo$_{2}$B$_{2}$C)
suggests that spin fluctuating could be the responsible mechanism for the
low temperature $\rho (T)$ behavior in this compound.\ Using the universal
relation for heavy fermion compounds; $A/\gamma ^{2}=1.0x10^{-5}\mu \Omega $
cm$(mole$K$)^{2}/mJ^{2}$\ \cite{Kadowaki} the resulted Sommerfeld
coefficient is $\gamma =89.4$\ $mJ/mol-$K$^{2}$ which is a low value
compared with that for PrNi$_{2}$B$_{2}$C ($211mJ/mol-$K$^{\mathbf{2}}$ \cite%
{Duran2PRB} by specific heat measurements), but enhanced value as compared
to the normal metal Co and to other borocarbides as YCo$_{2}$B$_{2}$C \cite%
{Massalami}, (Gd, Tb, Dy, Ho, Er, Tm)Ni$_{2}$B$_{2}$C \cite{MassalamiMagnon}%
, whose $\gamma $ is about $17$\ $mJ/mol-$K$^{2}$. \FRAME{ftbpFU}{3.4947in}{%
2.9758in}{0pt}{\Qcb{(Color online) The graph shows pressure effects on the $%
\protect\rho _{ab}(T)$\ for a PrNi$_{2}$B$_{2}$C single crystal up to 5.3
GPa. Vertical line indicates the increasing pressure. Insert is a zoom of $%
\protect\rho _{ab}(T,P)$\ at low temperature.}}{\Qlb{Fig4}}{fig4_falconi.eps%
}{\special{language "Scientific Word";type "GRAPHIC";maintain-aspect-ratio
TRUE;display "ICON";valid_file "F";width 3.4947in;height 2.9758in;depth
0pt;original-width 6.2759in;original-height 5.3385in;cropleft "0";croptop
"1";cropright "1";cropbottom "0";filename '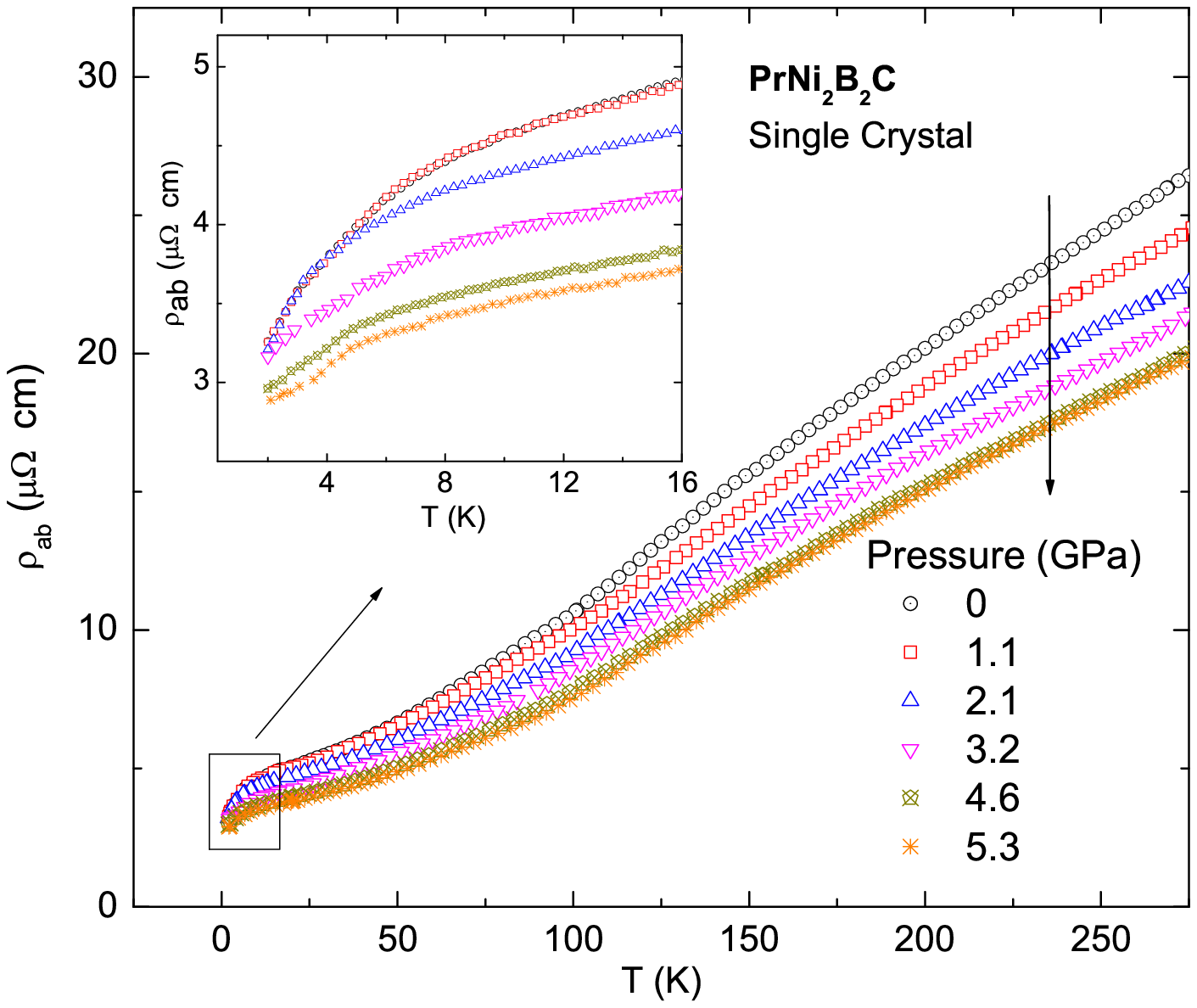';file-properties
"XNPEU";}}

\bigskip

Figs. 4 and 5 shown the $\rho _{ab}(T)$\ curves as a function of pressure
for PrNi$_{2}$B$_{2}$C and PrCo$_{2}$B$_{2}$C single crystals, respectively.
As we can see, these compounds reveal different pressure behaviors. The
overall trend of $\rho _{ab}(T)$ for the Pr-Ni based compound does not
change as the pressure increases up to 5.3 GPa. The linear behavior of $\rho
_{ab}(T)$ (extending from about 100 K to room temperature) is attributed to
electron-phonon scattering and under the applied pressures, it shows a slope
decreasing from 0.089 $\mu \Omega $cm/K to 0.069 $\mu \Omega $cm/K.
According to the inset of Fig. 4, the smooth drop of $\rho _{ab}(T)$ at low
temperature, which has been related to the decrease of magnetic scattering 
\cite{MiprimerPRB}, is reduced with applied pressures increasing up to 5.3
GPa. In the case of PrCo$_{2}$B$_{2}$C, as can be seen in panel \textbf{b}
of Fig. 5, the pressure effects are more stronger than in PrNi$_{2}$B$_{2}$%
C, mainly in the low temperature regime. From room pressure to about 1.7 GPa
the high temperature behavior of $\rho _{ab}(T),$ from 300 K to about 75 K,
remains without appreciable changes and with almost a constant slope of 0.30 
$\mu \Omega $cm/K. In an opposite way to PrNi$_{2}$B$_{2}$C, the low
temperature curvature of $\rho _{ab}(T)$, which is also associated with
magnetic correlations, tends strongly to be suppressed by pressure. This
tendency has also been observed by Massalami \textit{et. al.} \cite%
{MassalammiJMMM2004} applying pressures up to about 1.2 GPa, the maximum
pressure value they applied. Similar to their results, we also observed that
the cuadratic behavior of $\rho _{ab}(T)$ at low temperatures is maintained
under 1.2 GPa. However, we found, applying pressures higher than 1.7 GPa on
this compound, a distinctive characteristic, namely the change of the low
temperature curvature from concave to convex and the complete disappearance
of the resistivity drop above 2.9 GPa (see panel \textbf{b} in Fig.5).
Interestingly, at this pressure the \textit{T}$^{2}$ behavior disappeared
and instead there is an appearing of a type plateau zone in $\rho _{ab}(T)$
which start about 15 K and extend down to 1.8 K, the lowest temperature
available in our experiments. Increasing the pressure up to 4.4 GPa, this
zone of constant resistivity is extended from 1.8 K up to 20 K. At this
pressure, the overall high temperature behavior of $\rho _{ab}(T)$ remains
almost with the same slope of about 0.30 $\mu \Omega $cm/K.\ 

The above experimental facts reveal that the electronic properties of PrCo$%
_{2}$B$_{2}$C are more pressure sensitive than that for PrNi$_{2}$B$_{2}$C,
mainly at the low temperature regime. Interestingly, we note that the shape
of $\rho _{ab}(T)$ for PrCo$_{2}$B$_{2}$C under pressure tends to be
qualitatively similar to that for nonmagnetic YCo$_{2}$B$_{2}$C at room
pressure \cite{Massalami}. Thus, it seems the effect of pressure in this
compound is to suppress the magnetic correlations which originate the low
temperature scattering behavior.\FRAME{ftbpFU}{3.2638in}{3.6227in}{0pt}{\Qcb{%
(Color online) In a) it is presented $\protect\rho _{ab}(T)$\ for PrCo$_{2}$B%
$_{2}$C single crystal under several pressures up to 4.4 GPa. Panel b) shows
a view of $\protect\rho _{ab}(T)$\ at low temperatures and for different
pressure values.}}{\Qlb{Fig5}}{fig5_falconi.eps}{\special{language
"Scientific Word";type "GRAPHIC";maintain-aspect-ratio TRUE;display
"ICON";valid_file "F";width 3.2638in;height 3.6227in;depth
0pt;original-width 6.3131in;original-height 7.0119in;cropleft "0";croptop
"1";cropright "1";cropbottom "0";filename '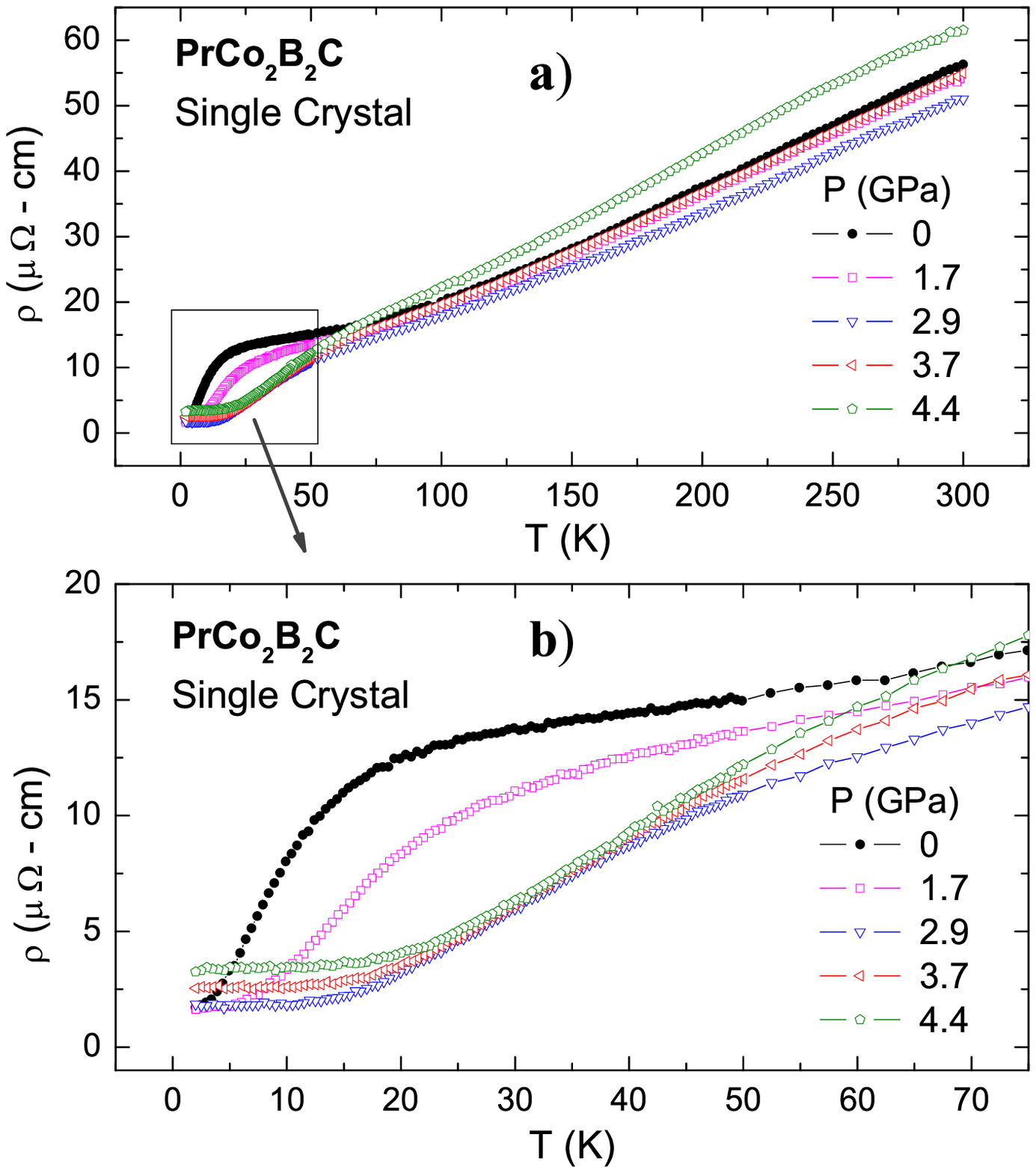';file-properties
"XNPEU";}}

According to Fig 5, the resistivity can be fitted to a $T^{2}$ behavior only
for a region of relatively low temperatures, interestingly under pressure
the fitting range is extended. As already we pointed out, at atmospheric
pressure the fitting goes from 2 up to about 8 K with a cuadratic
coefficient \textit{A}, equal to 0.08 $\mu \Omega $cm/K$^{2}$. At 1.7 GPa
the interval extends from 2 to about 19.5 K, with a value of \textit{A}
decreased to 0.018 $\mu \Omega $cm/K$^{2}$. Once the applied pressure
reaches the value of 2.9 GPa, it was not more possible to fit a cuadratic
function to the curvature of $\rho _{ab}(T)$ because the like plateau zone,
also presented for the curves at 3.7 and 4.4 GPa. This $\rho _{ab}(T)$
behavior at low temperature and pressures is accounted by the spin
fluctuation scenario \cite{Rossiter}, which also take in account the
decreasing of the \textit{A} parameter with the applied pressure.
Additionally, relatively low magnetic fields decreases the \textit{A}
parameter in a linear form in PrCo$_{2}$B$_{2}$C \cite{Morales}, which has
been claimed to be due to quenching of the spin fluctuation. Similar results
have been found in other systems as for example, Ce$_{0.8}$(Pr, Nd)Ni$_{5}$,
indicating that spin fluctuations tends to be suppressed both by pressure
and magnetic fields \cite{Marian,Willis}. The microscopic character of the
state resulting from applying a magnetic field is completely different from
that obtained with applying pressure; the first remains magnetic, whereas
the last tends to be a real non-magnetic state; one where no microscopic
magnetic moments exists. This is important because in the spin fluctuations
model the state above $T_{sf}$ (where the spin fluctuation appears and which
coincide with the temperature below which a $T^{2}$ law in resistivity is
valid) is non the non magnetic state (like in the stoner model) but a
magnetic state, where local moments still exist, but long range order tend
to be destroyed by the fluctuations.

The collective modes described by the spin fluctuations can readily be
excited at relatively low temperature, where the stoner excitations are
still very small but we assume they can be suppressed by two factors:
intense magnetic field and/or external applied pressure. Pressure increases
the correlation there exist between \textit{f} ions and promotes the
itinerance of \textit{f} electrons. As a result the \textit{f}-density of
state near the Fermi level is lowered modifying the electron structure and
influencing thus the prevailing long range order between the band electrons.%
\FRAME{ftbpFU}{3.4177in}{2.7882in}{0pt}{\Qcb{(Color online) Normalized
R(T)/R(40 K) curves at several pressures up to 2.5 GPa for polycrystalline
PrNi$_{2}$B$_{2}$C. Insert shows the behavior of $\protect\rho (P)$\ at 260
K.}}{\Qlb{Fig6}}{fig6_falconi.eps}{\special{language "Scientific Word";type
"GRAPHIC";maintain-aspect-ratio TRUE;display "ICON";valid_file "F";width
3.4177in;height 2.7882in;depth 0pt;original-width 6.3157in;original-height
5.1448in;cropleft "0";croptop "1";cropright "1";cropbottom "0";filename
'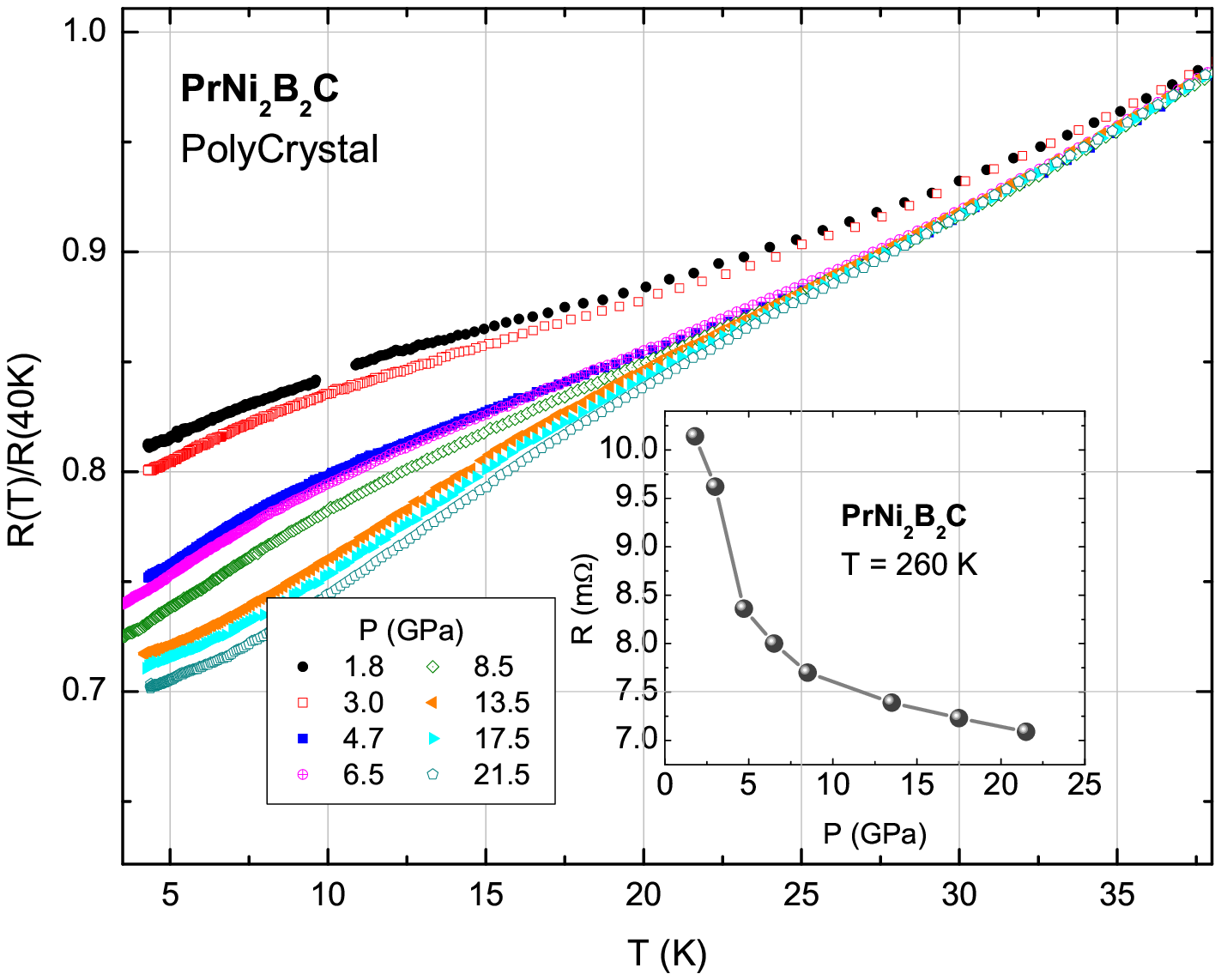';file-properties "XNPEU";}}

In order to know if the $\rho (T)$ of PrNi$_{2}$B$_{2}$C could follow a
similar behavior to that for its isomorphs PrCo$_{2}$B$_{2}$C at higher
pressures, we carried out measurements for polycrystalline sample at several
pressures up to 21.5 GPa. Interestingly, we found there is a marked tendency
of $\rho (T)$ at low temperature to behave similar to that for PrCo$_{2}$B$%
_{2}$C (see Fig. 6). At about 13.5 GPa there is a change from negative to
positive curvature of $\rho (T)$\ at temperatures lower than about 15 K.
This curvature change was also found in Pr-Co system but at about 1.7 GPa.
It is important to mention that from the $\rho _{ab}(T=260K,P)$ curves for
PrNi$_{2}$B$_{2}$C (see inset of Fig. 6), and PrCo$_{2}$B$_{2}$C (not
shown), we discarded some structural phase changes that could be related to
these effects. The fact that Pr-Ni system requires more pressure to behaves
almost in the same way to Pr-Co system at low temperature, could be related
to changes in the \textit{c- }parameter.\FRAME{ftbpFU}{3.3114in}{2.5365in}{%
0pt}{\Qcb{(Color online) $R(T)$ curves for polycrystalline PrPt$_{2}$B$_{2}$%
C, measured up to 21.5 GPa. High pressure tends to destruct
superconductivity. Insert shows the low temperature behavior of $R(T).$}}{%
\Qlb{Fig7}}{fig7_falconi.eps}{\special{language "Scientific Word";type
"GRAPHIC";maintain-aspect-ratio TRUE;display "ICON";valid_file "F";width
3.3114in;height 2.5365in;depth 0pt;original-width 10.9624in;original-height
8.3835in;cropleft "0";croptop "1";cropright "1";cropbottom "0";filename
'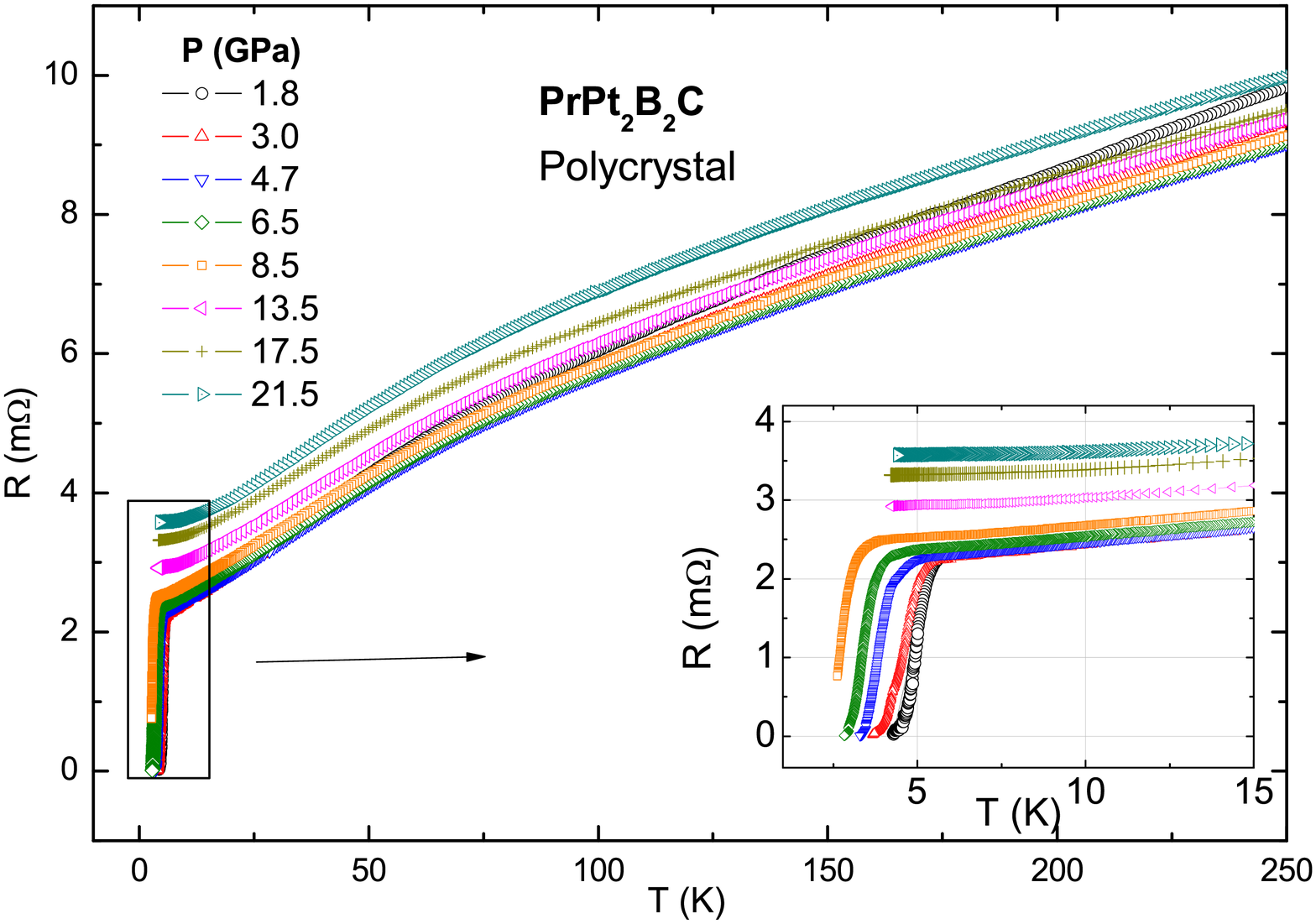';file-properties "XNPEU";}} The main difference in the
unit cell of these two compounds stands from this parameter, which is
biggest in Pr-Co system and related to the modifiable tetrahedral B-T-B
angle. On the other hand, it is known that for the spin fluctuator YMn$_{2}$%
, the existence of a magnetic moment on Mn sites depends largely on the
interatomic Mn-Mn distance \cite{Hauser}. Above a critical distance there
exists a magnetic moment.\ Such distance plays a key-role in determining the
magnetic properties and is sensitive to external or internal perturbations.
The case for Pt in PrPt$_{2}$B$_{2}$C could be similar. As we already point
out this compound shows an upward curvature in $\rho (T)$\ at high
temperature which has been related to crystalline electric field (CEF)
effects \cite{Duran2PRB}. We make resistivity measurements for
polycrystalline PrPt$_{2}$B$_{2}$C under several pressures up to 21.5GPa
(see fig. 7). As it can be observed, the negative curvature of $\rho (T)$\
at high temperature is not appreciably modified under pressure and the main
changes are at low temperatures. The superconducting transition temperature,
T$c$, decreases at the rate dT$c$/dP = -0.34 \ K/ GPa (see Fig. 8). It seems
there is not correlation with the decreasing of T$c$ and the unmodified
curvature related to CEF effects. A positive magnetoresistance at low
temperature in this compound has been associated with spin fluctuation \cite%
{Duran2PRB}, however although this picture follows the same trends of
Pr(Ni,Co)$_{2}$B$_{2}$C, further investigations are inquired in order to
clarify this matter.\FRAME{ftbpFU}{3.5751in}{2.9845in}{0pt}{\Qcb{(Color
online) Decreasing of the superconducting transition temperature for PrPt$%
_{2}$B$_{2}$C, as function of pressure. The rate of decreasing of the
transition looks normal for a \textit{d} electronic compound.}}{\Qlb{Fig8}}{%
fig8_falconi.eps}{\special{language "Scientific Word";type
"GRAPHIC";maintain-aspect-ratio TRUE;display "ICON";valid_file "F";width
3.5751in;height 2.9845in;depth 0pt;original-width 8.3083in;original-height
6.9272in;cropleft "0";croptop "1";cropright "1";cropbottom "0";filename
'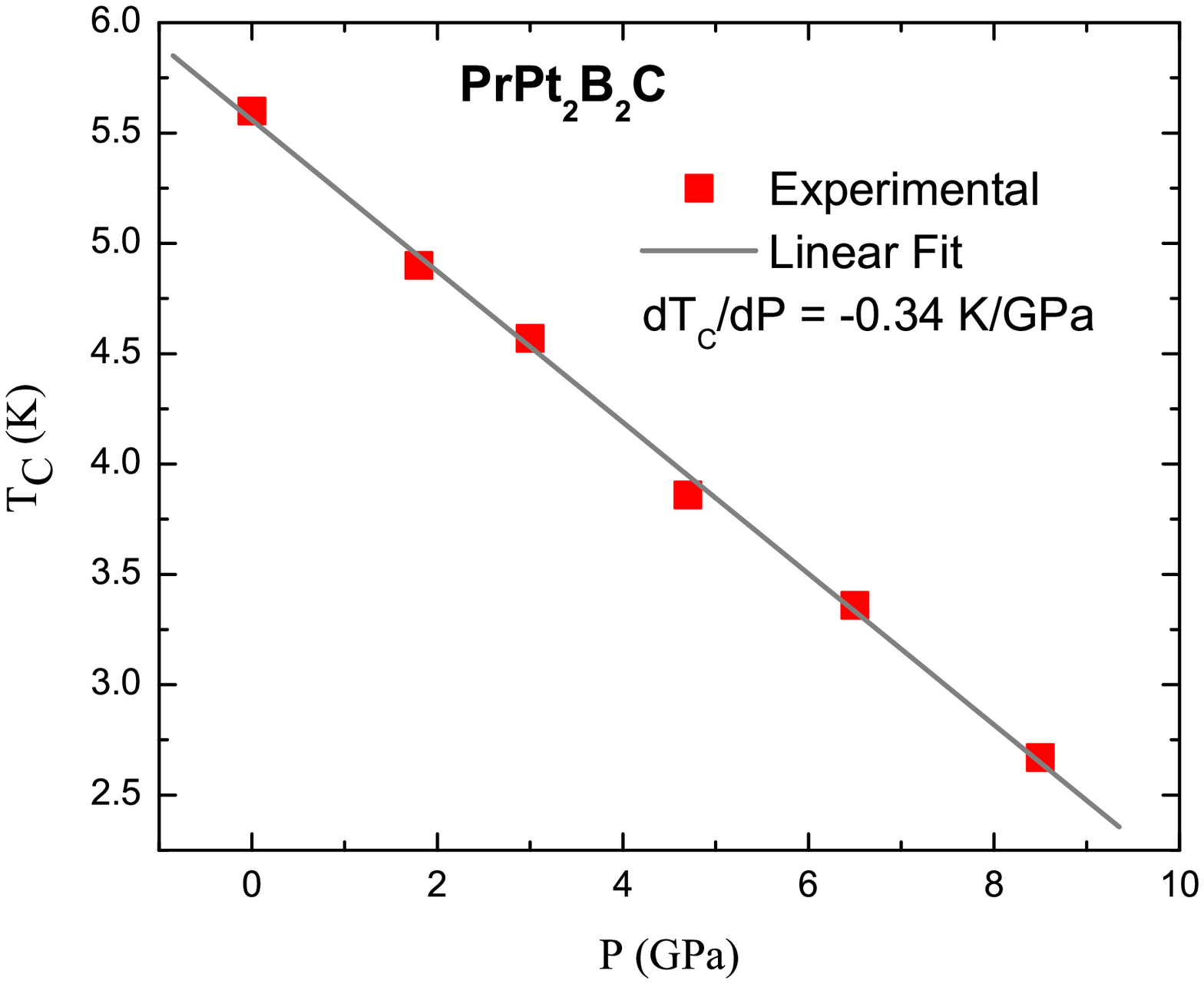';file-properties "XNPEU";}}

\section{Conclusions}

High pressure resistivity measurements in Pr(Co,Ni,Pt)$_{2}$B$_{2}$C has
been made. The first and foremost fact we found is that applied pressures of
about 4.0 GPa are able to change drastically the low temperature resistivity
behavior of PrCo$_{2}$B$_{2}$C, but it requires $\sim $ 13.0 GPa in order to
attain similar changes for PrNi$_{2}$B$_{2}$C. This means that the low
temperature electronic transport properties of PrCo$_{2}$B$_{2}$C are more
pressure sensitive than that for the isomorphous PrNi$_{2}$B$_{2}$C.
Evidence for spin fluctuation in PrCo$_{2}$B$_{2}$C is reported from the
cuadratic behavior of its resistivity at low temperature, and from the
decreasing of the cuadratic coefficient as a function of pressure. For\ PrCo$%
_{2}$B$_{2}$C, the magnetic scattering related to spin disorder is
suppressed at 2.9 GPa, but it remained observable at less under 5.3 GPa for
PrNi$_{2}$B$_{2}$C. For the case of the superconductor PrPt$_{2}$B$_{2}$C,
pressure does not modify the $\rho (T)$ curvature related to CEF effects,
but decreases Tc at the rate dTc/dP = -0.34 K/GPa. Finally, although these
conclusions are no decisive, we believe they would stimulate further
experimental and theoretical studies for a better understanding of the
pressure effects in the \textit{R}T$_{2}$B$_{2}$C compounds, which is far
for complete.

The authors acknowledge the MCBT, Institut N\'{e}el, (CNRS) \& UJ for the
time granted to perform some high pressure experiments. R. F. thanks to
SEP-PROMEP, UJATAB-CA175 for support. R. E. thanks UNAM-DGAPA, project No
IN-101107. We thank to S. Bern\`{e}s for the crystallographic studies and F.
Silvar for liquid He supply.

\end{document}